\documentstyle[aps,prb]{revtex}

\begin{document}
%\draft
\twocolumn[
\title{ Symmetry in the insulator -  quantum Hall - insulator transitions 
observed in a Ge/SiGe quantum well }
\author{
M. Hilke, D. Shahar, S.H. Song, D.C. Tsui }

\address{ Department of electrical engineering, 
Princeton University, Princeton, NJ-08544}
\author{
Y.H. Xie and Don Monroe }

\address{ Bell Laboratories, Lucent Technologies, 
Murray Hill, NJ-07974}

\date{July 7, 1997}

%\maketitle
%\begin{center}
%\abstract{
%\date{
\begin{center}
\parbox{15cm}{\small 
We examine the magnetic field driven insulator-quantum Hall-insulator 
transitions of the two dimensional hole gas in a Ge/SiGe 
quantum well. We observe direct transitions between low and high 
magnetic field insulators 
and the $\nu=1$ quantum Hall state. 
With increasing magnetic field, the transitions from 
insulating to quantum Hall and quantum Hall to insulating are 
very similar with respect to their transport properties.
We address the temperature dependence around the transitions and show  
that the characteristic energy scale for the high field 
transition is larger.

 }
\end{center}

\maketitle
]

The quantum Hall (QH) effect is an excellent system to study transitions 
between insulating and metallic behaviors in two dimensions. In the integer 
QH effect the basic physics is governed by the Landau 
levels (LL). The system becomes  insulating at $T=0$ when the energy of the 
lowest LL crosses and exceeds the Fermi energy. In this case we have 
a transition from the QH state, characterized by the filling factor $\nu=1$, 
to an insulator. This was  observed by 
Paalanen et al. \cite{palaanen} in low mobility samples. For much cleaner 
samples with very high mobilities,  
the transition to an insulating behavior can occur from  
much lower fractional QH states like $\nu=1/5$ as observed  by 
Jiang et al. \cite{jiang2}. This transition was interpreted as 
evidence  for the 
formation of a Wigner crystal. In the  zero magnetic field case a similar  
transition 
was observed by Kravchenko et al. \cite{kravchenko} but 
as a function of density.

Another intriguing  phenomenon has been observed recently 
by Jiang et al. \cite{jiang} and others \cite{hughes,wang}: they observed a 
direct transition from a low magnetic field (B) insulating phase 
to the $\nu=2$ state.
Later Shahar et al. \cite{shahar} observed a transition to 
 the $\nu=1$ state in a 2 dimensional electron system 
and Song et al. \cite{song}
saw direct transitions between a low B insulator and the  
$\nu=3$, $\nu=2$ and $\nu=1$ quantum Hall states, depending on density.
The general framework for understanding 
 these insulator-to-QH states 
transitions can be found in the pioneering work of 
Kivelson, Lee and 
Zhang (KLZ) \cite{klz} in terms of a global phase diagram (GPD). 
KLZ's theory is successful in explaining  
transitions in the integer and fractional QH regime but it fails 
in accounting for 
the direct transitions to higher order QH states such as the 
$\nu=3$ state 
\cite{song}. It is therefore essential  to study  
these transitions in detail in order to improve our understanding 
of the GPD.

The purpose of this work is to give a better understanding of the 
insulator-QH-insulator transitions
 by concentrating on transitions involving only the  $\nu=1$ QH state. 
In particular the similarities 
between the low and high B-field  transitions will be demonstrated.
We will further present a detailed  study of the temperature (T) 
dependence of the resistivity around both, 
the low-B field transition, which  separates  
a low B insulator from the spin split $\nu=1$ QH state, and 
the high B-field transition that separates the QH state $\nu=1$ from 
the insulating high B-field phase.

This paper is organized as follows. After discussing 
the experimental details we will focus on the diagonal resistivity, 
which allows us to identify the 
low and high B-field
transition points . We study the 
T-dependence of the slopes of the diagonal resistivity 
at the transitions. 
Our main result is that there is a clear  similarity 
between both transitions,  essentially differing only by their
energy scales.
We conclude with a discussion of our 
results.

The results of this work were obtained from a 2 dimensional 
 hole system (2DHS) in 
a strained Ge layer. The 2DHS is contained in the Ge layer which is under 
compressive strain. The sample studied was grown by MBE technique and 
consists of a graded buffer $\mbox{Si}_{1-x}\mbox{Ge}_x$ layer grown 
on a Si (100) substrate, followed by a uniform buffer 
$\mbox{Si}_{0.4}\mbox{Ge}_{0.6}$ layer and a $150$ \AA{ }thick Ge layer 
sandwiched in between  $\mbox{Si}_{0.4}\mbox{Ge}_{0.6}$ layers where 
Boron modulation doping is placed. The 2DHS has a mass of $\sim$0.1 $m_e$, 
which is density dependent.
\cite{song,xie}

We could vary the density and mobility by 
means of a metallic front gate. A MOSFET structure was made by depositing 
an insulating layer between the metal and the cap layer. A standard 
Hall bar was processed with Ti/Al gate and Al/Au alloyed Ohmic 
contacts with a 50 $\mu$m wide channel and 600 $\mu$m apart voltage 
probes. By applying a gate voltage between 0 V and 6 V we could vary 
the density between $n=0.7-6\times 10^{11} \mbox{ cm}^{-2}$ and the mobility 
between $0.3-20\times 10^3 \mbox{ cm}^2/Vs$. In our whole  gate 
voltage range we observed no gate leakage.

The measurements were performed in a He-3 refrigerator at T's
ranging from 300 mK to 7 K, using an AC lock-in technique with 
an excitation current of 0.2 nA. The results 
were reproducible with a current of 0.05 nA, within experimental 
accuracy. DC measurements were performed to 
check for consistency. The results in this work are obtained 
at a fixed gate voltage of $V_G=5.2$ V. The measured density was 
$0.87\times 10^{11} \mbox{ cm}^{-2}$.

In fig.\ 1 we show the diagonal resistivity, $\rho_{xx}$,
as a function of B, for various T's
ranging from 0.3 K to 4.2 K. The Hall resistivity is plotted at 1.8 K. 
Three regions can clearly be 
distinguished, which we now describe in order of increasing 
B.  

\input epsf
\begin{figure}
%\begin{center}
%\leavemode
%hbox{%
\epsfysize=7cm

\epsfbox{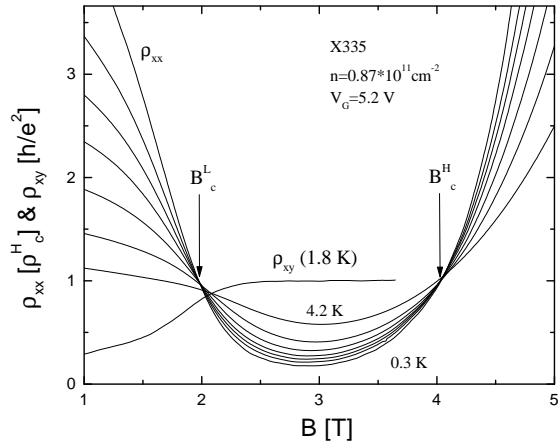}
%\end{center}
\caption{Diagonal and Hall resistivity as a function of magnetic field. 
$\rho_c^H=2.2 h/e^2$. The temperatures are: 
0.3, 0.55 , 0.75 , 0.9 , 1.2, 2.4 and 4.2 K}
\end{figure}

The first region is a low B-field insulating phase characterized 
by an increasing resistivity with decreasing T and a linear  
Hall resistivity as B tends to zero. 
The second region is the $\nu=1$ QH state. 
At  $B_c^L$  the 
two regions are clearly separated by a T-independent $\rho_{xx}$ (at low 
temperatures).
The $\nu=1$ QH state is characterized by a well-developed 
 plateau in the Hall resistivity, with the expected value of $\rho_{xy}=h/e^2$,
and a decreasing $\rho_{xx}$ with decreasing T. 
Finally, the last region is similar to the first and is also 
characterized by a diverging diagonal resistivity 
when decreasing T. 
Here again we have a well-defined 
transition point, with a T-independent resistivity, 
$\rho_c^H=\rho_{xx}(B_c^H)$, 
for low enough temperatures.

The transition points can be reliably extracted by   plotting, as we do 
in fig.\ 2,    
the resistivities as a function of T for different 
values of B. In fig.\ 2 a) we present 
the high B-field transition and observe that the  plot corresponding 
to the magnetic field $B_c^H=4.04$ T has almost no T-dependence. The 
resistivity remains constant within 1 \% between 0.3 K and 3.2 K.
For higher magnetic fields the resistivity diverges with 
decreasing T whereas for lower magnetic fields the dependence is
opposite. The low B-field transition, presented in fig.\ 2 a), 
is very similar.
Extracting the magnetic field 
corresponding to the
T-independent behavior below 1.8 K, we obtain a critical B for the 
high B-field transition of $B_c^L=1.975$ T.

\input epsf
\begin{figure}
%\begin{center}
%\leavemode
%hbox{%
\epsfysize=7cm

\epsfbox{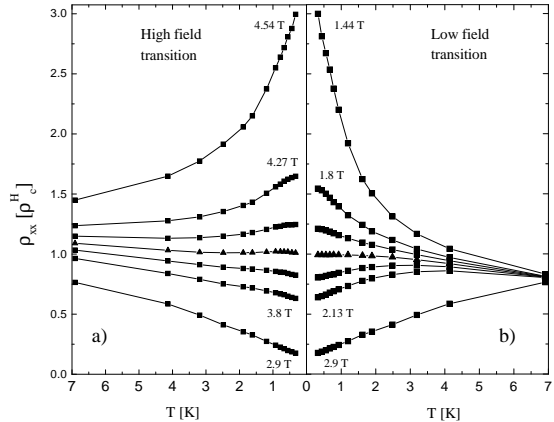}
%\end{center}
\caption{Temperature dependence of the resistivities around the low and 
high field transitions. In fig.\ 2 a) the magnetic fields corresponding 
to the central resistivity curves are 3.94, 4.04 and 4.14 T and in 
fig.\ 2 b) they are 2.05, 1.975 and 1.9 T.}
%\end{center}
\end{figure}

Since the main idea in this work is to concentrate on properties related 
to transitions into  insulating phases we chose a system with  a strong 
T-dependence in these phases, in particular in the low B-field phase. This 
implies the choice of a system with strong disorder, which 
leads to a high value of the minimum diagonal  
resistivity in the QH state, even at 0.3 K. Therefore, we are unable to  
directly compare our result $\rho_c^H=2.2 h/e^2$ with the value $h/e^2$ 
obtained in the 
work of Shahar et al. \cite{shahar3}, 
which demonstrated  the universality of the resistivity at $B_c^H$.
They had only included in their study samples for which the  minimum diagonal  
resistivity in the 
QH state was vanishing at low T. In fact, when we lower the gate voltage, 
i.e., reduce the effective disorder strength, the value of $\rho_c^H$ 
approaches $h/e^2$.\cite{song}

We can now turn to the main result of this work, concerning
the similarities between 
both transitions. Extracting the resistivities from fig.\ 2, 
we first note that 
\begin{equation}
\rho_c^L=\rho_c^H\pm 3 \%,
\end{equation} 
where $\rho_c^L=\rho_{xx}(B_c^L)$.
It is interesting to mention that 
a similar approximate relation (1) 
holds for  the $\nu=2$ 
QH state to high and low B-field transitions \cite{jiang,hughes,wang}.
These experiments were performed in systems where the $\nu=1$ state 
was not resolved. 
Since relation (1) applies only at the transition points, 
we will now concentrate on the behavior around the transition points and 
study the similarity between both transitions 
as a function of B and T. We will first focus on the B-dependence 
and then turn to the T-dependence. 

The best way to compare the B-dependence 
is to overlap both transitions as a function of the filling factor 
$\nu$, which is dimensionless. 
We therefore convert our diagonal resistivity data of 
fig.\ 1 as a function of $\nu$, which is  
obtained by measuring the density from the Hall resistance. 
The result is represented in fig.\ 3. We note that the two transitions are 
almost indistinguishable. The main difference is the effective  temperature 
of the two  transitions, which can be associated to a 
characteristic energy scale of each transition. The high value of 
$\nu_c^H=0.87$ is consistent with a transition that occurs at a high 
disorder value as expected accordingly to 
KLZ's \cite{klz} phase diagram.

\input epsf
\begin{figure}
%\begin{center}
%\leavemode
%hbox{%
\epsfysize=6cm

\epsfbox{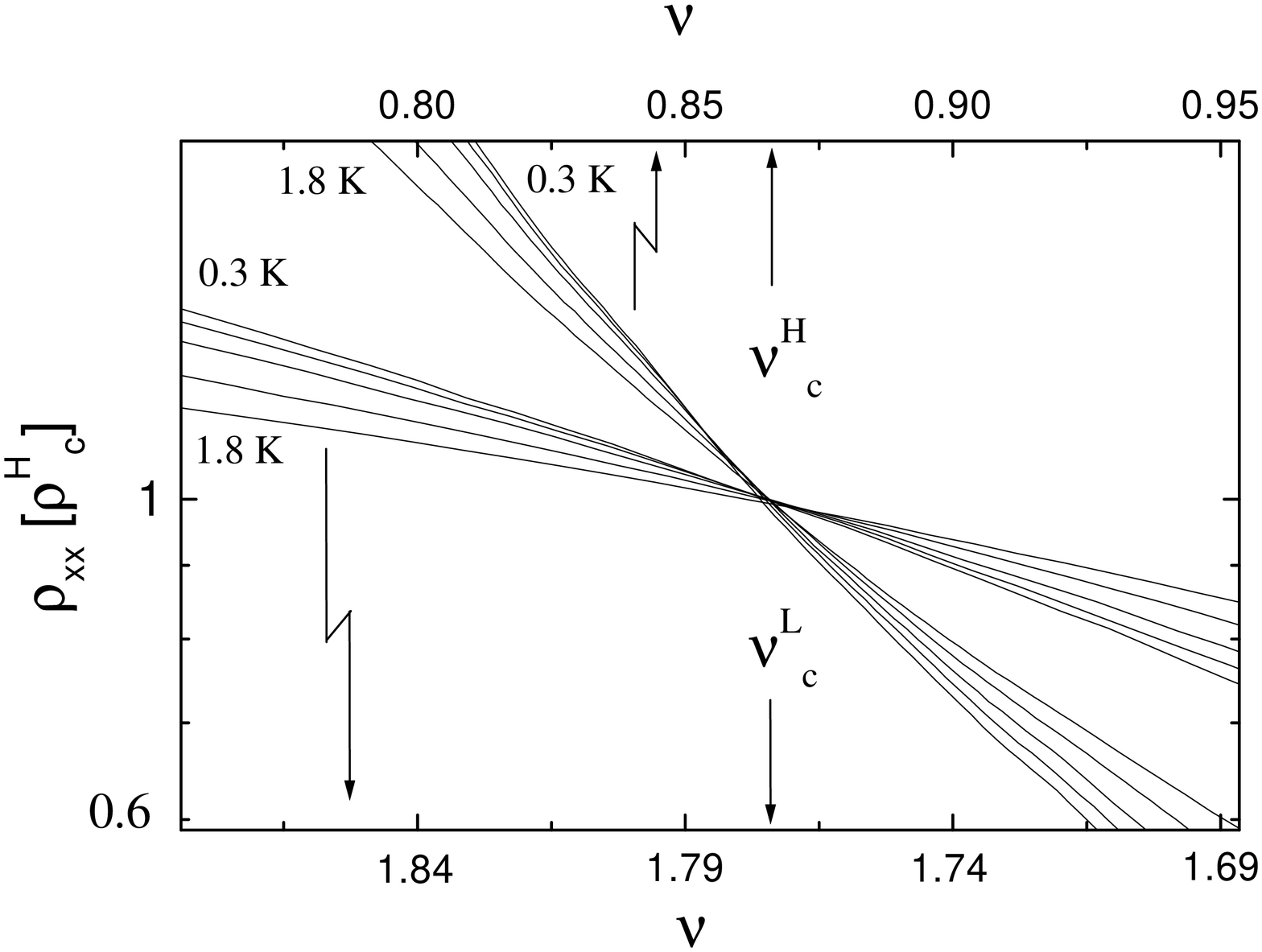}
%\end{center}
\caption{Resistivities, on a Log scale, as a function of filling factor. 
The group of curves with  steeper slopes corresponds to the high B-field 
transition with critical filling factor $\nu_c^H=0.87$ obtained from  
 $B_c^H$. The critical filling factor $\nu_c^L=1.77$ of the low B-field 
transition corresponds to $B_c^L$. The bottom $\nu$ scale is reversed.}
\end{figure}

In the following we present our temperature dependence analysis at the 
transitions.
There are several ways for studying this temperature dependence. 
The prevailing studies \cite{jiang,hughes}  assume a scaling behavior around 
the 
critical point and try to collapse all data on one curve. Because 
this study assumes  a scaling 
behavior  it is difficult to extract any information 
other than a scaling exponent. We will use an  
approach which is more general in the sense that 
it does not suppose any scaling form. Following Wang et al. \cite{wang}, 
this approach consists 
simply of taking the slopes of the diagonal resistivity at the 
transition, i.e., near the T-independent point. In order 
to obtain a dimensionless result we evaluate 
$\alpha_0^{-1}
=\left(\frac{\partial \log(\rho_{xx})}{\partial\nu}\right)|_{\nu_c}$, which 
is simply the slope of the plots in fig.\ 3. In addition to giving a 
dimensionless result,
this method has the advantage that $\log(\rho_{xx})$ 
can be well approximated by a linear function of $\nu$, as was demonstrated 
by Shahar et al. \cite{shahar4}, for the high B-field transition. A new 
result of this work is that for the 
low B-field transition 
we also observe a similar linear dependence. 
This linear dependence of the slopes allows us to extract 
 the slopes in a  well defined and straightforward way  for 
the different temperatures. It is interesting to note that a similar linear 
dependence is also observed as a function of density in the zero B 
metal-insulator transition.\cite{simonian} 

In fig.\ 4 we plot the dimensionless 
inverse-slopes, $\alpha_0$, on a Log-Log scale. It is evident from 
the data that it is 
not possible to extract any power law consistently for neither the low nor 
the high 
B-field transitions.

\input epsf
\begin{figure}
%\begin{center}
%\leavemode
%hbox{%
\epsfysize=6cm

\epsfbox{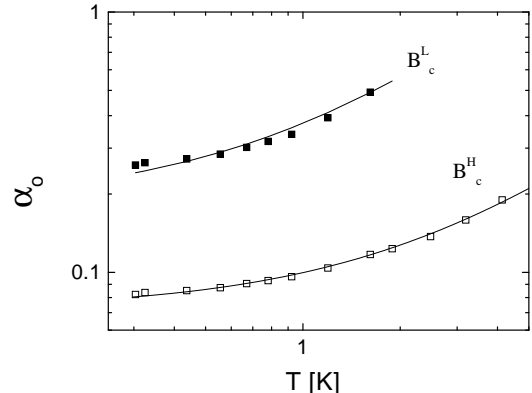}
%\end{center}
\caption{$\alpha_0$ for the low B-field transition (solid symbols) and the 
high B-field transition (open symbolss). The lines are the fits obtained 
from fig.\ 5. }
\end{figure}
It is possible, however, to 
extract some interesting information from fig.\ 4 concerning 
the different underlying energy scales.
When the slopes of the low B-field transition are 
divided by 3 we can overlap the low B-field transition 
and the high B-field transition for the lowest temperatures.
This is consistent 
with a lower characteristic energy scale for the low B-field transition.

We further tried to fit the T-dependence following Shahar et al. 
\cite{shahar4} The suggestion is that $\alpha_0$ 
follows a linear T-dependence. We observe in fig.\ 5 a) a 
 good agreement with this behavior for the high field transition. In the 
low field transition 
 case, fig.\ 5 b), 
the data starts to deviate from this behavior at 1.7 K. This 
deviation can also be understood in terms of the lower effective temperature 
scale in the low field transition,
which drives the system faster into 
a different behavior. The same ratio 3 between both transitions 
is obtained for the zero-temperature 
extrapolated value of $\alpha_0$, but a ratio 
of 6 for the linear temperature coefficient.

\input epsf
\begin{figure}
%\begin{center}
%\leavemode
%hbox{%
\epsfysize=6cm

\epsfbox{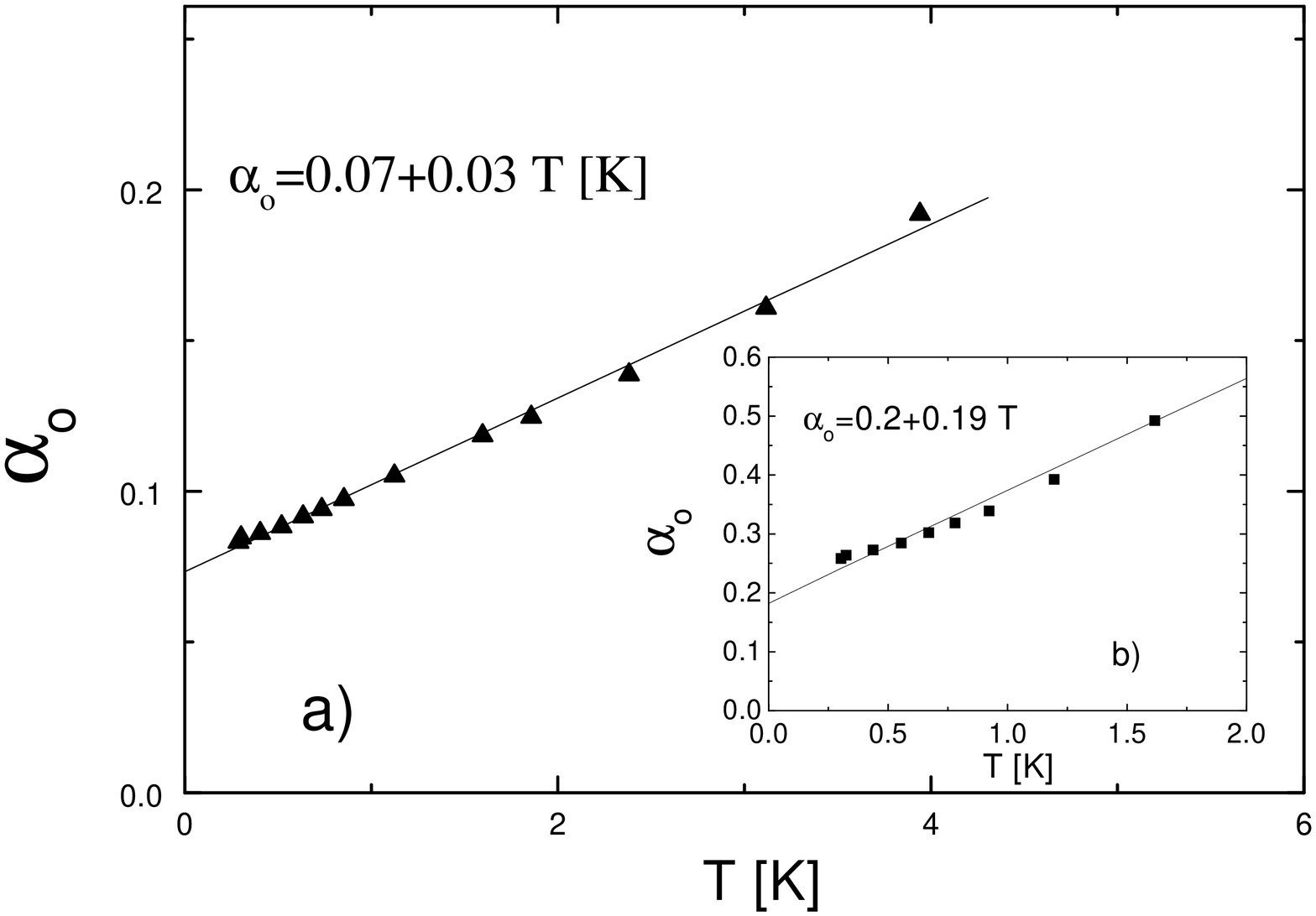}
%\end{center}
\caption{$\alpha_0$ on a linear graph as a function of temperature for the high field 
transition. The inset shows the low field transition up to 1.7 K. The 
straight lines are linear fits to the data.}
\end{figure}

Coming back to fig.\ 2, similar conclusions can be drawn, concerning 
the different energy scales. 
For the 
high B-field transition, $\rho_{xx}$ remains constant within 1 \% over 
a wide temperature range, i.e., from our base temperature, T=0.3 K,  
to approximately 3.2 K, whereas for the low 
B-field transition $\rho_{xx}$ remains constant only up to 1.7 K. 
At 7 K the relative deviation to the 0.3 K value is 7 \% for the 
high B-field transition but 20 \% for the low B-field transition. 
We here note a different qualitative behavior: at the low B-field transition  
$\rho_{xx}$ {\em decreases} with higher temperatures whereas at the high 
B-field 
transition $\rho_{xx}$ {\em increases} with higher temperature.

In the following we discuss our results in light of  existing theories.
Theories of disordered systems predict overall localized states in 
two dimensions 
and at zero B.\cite{abrahams} However, 
when the quantization  due to LL becomes important at high fields it 
is expected that extended states exist at the center of LL's.\cite{janssen}

There are several theoretical and numerical results 
dealing with the crossover from the  localized zero B-field state to the 
delocalized high B-field state. The original argument of  
Khmel'nitzkii and Laughlin \cite{khmelinski} 
describes the crossover 
as follows: the energies of the  
extended states originally at the center of a Landau level at high fields, 
 float up with decreasing magnetic field. This is commonly referred 
to as the  floating up picture. When the magnetic 
field is decreased the Fermi energy  crosses the up-floating 
lowest energy-level 
of extended states and yields an insulator. In this framework, the 
low field transition is also induced by the crossing of a LL, 
in analogy to the high field transition. This picture has gained 
recent support by numerical and theoretical results.\cite{kunyang}
Pruisken \cite{pruisken} describes the crossover in terms of a field 
theory, recovering the localized case as the simplectic limit and the 
delocalized case as the unitary limit.
An alternate description  is given by Liu et al.\cite{liu} 
Their numerical results 
suggest that in the center of each LL there is a localization-delocalization 
transition with increasing B field. 

The similarity between both transitions seems to favor the floating up 
picture as they would both be described by the LL level crossing of the 
Fermi energy. However, the floating picture does not predict a 
substantial different energy scale for both transitions, as opposed 
to the inter-LL localization-delocalization picture. All theories 
described above assume a single-particle picture, but at low fields
other energy scales like  
interparticle interactions become relatively more important and 
could alter the simple one-particle physics.

In conclusion, we analyzed the insulator - quantum Hall state $\nu=1$ - 
insulator 
transitions. The transitions are 
characterized by a temperature independent diagonal resistivity.
The main result arising 
from this study is that both transitions are very similar in terms of 
the magnetic field and temperature dependence so that it seems 
reasonable to assume a similar mechanism for both transitions.

This work was supported by the National Science Foundation and 
M.H. was supported by the Swiss National Science Foundation.

\end{document}